\input{epsf}
\documentstyle[eqsecnum,multicol,aps]{revtex} 
\renewcommand{\narrowtext}{\begin{multicols}{2} 
\global\columnwidth20.5pc}
\renewcommand{\widetext}{\end{multicols} \global\columnwidth42.5pc}
\begin{document}
\draft
\preprint{HEP/123-qed}
\title{X-ray anomalous scattering investigations on the charge order
in $\alpha^\prime$-NaV$_2$O$_5$}
\author{S.~Grenier${^1}$, A.~Toader$^{1,*}$, J.E.~Lorenzo$^{1,2}$, 
Y.~Joly${^1}$, B.~Grenier${^3}$, S.~Ravy${^4}$, L.P.~Regnault${^3}$, \\
H.~Renevier${^1}$, J.Y.~Henry${^3}$, J.~Jegoudez${^5}$ and A.~Revcolevschi${^5}$}
\address{
         ${^1}$Laboratoire de Cristallographie, CNRS, BP 166, 38042
Grenoble cedex 9, France\\
         ${^2}$European Synchrotron Radiation Facility, Av.~Jules~Horowitz, 
BP 220, 38042 Grenoble cedex 9, France\\
         ${^3}$Commissariat \`{a} l'Energie Atomique,
D\'{e}partement de Recherche Fondamentale sur la Mati\`{e}re Condens\'{e}e,
SPSMS/MDN, 38054 Grenoble cedex 9, France\\
         ${^4}$Laboratoire de Physique des Solides, CNRS,
Universit\'{e} Paris-Sud, B\^{a}t 510, 91405 Orsay cedex, France\\
         ${^5}$Laboratoire de Physico Chimie des Solides, CNRS, UMR 8648,
Universit\'{e} Paris-Sud, B\^{a}t 414, 91405 Orsay cedex, France\\
}
\date{\today}
\maketitle
\begin{abstract}
Anomalous x-ray diffraction studies show that the charge ordering in $\alpha^\prime$-NaV$_2$O$_5$
is of zig-zag type in all vanadium ladders. We have found that there are two models of the 
stacking of layers along \emph{c-}direction, each of them consisting of 2 degenerated patterns, and that 
the experimental data is well reproduced if the 2 patterns appears simultaneously. We believe that the low
temperature structure contains stacking faults separating regions corresponding to the four possible patterns.
\end{abstract}
\narrowtext

In the recent past years $\alpha^\prime$-NaV$_{2}$O$_{5}$ has raised a great deal of interest as being the 
second inorganic compound, after CuGeO$_{3}$, where a quantum antiferromagnetic state is achieved at 
the expense of a lattice distortion\cite{Isobe96}. At  T$_{C}$=34K, it exhibits a 
structural\cite{Tapan98} as well as a magnetic phase transition that lifts the magnetic degeneracy, 
opening an energy gap ($\Delta$=9.8 meV) between the ground state, a non-magnetic spin-singlet, and 
the lowest magnetic spin-triplet\cite{Yosihama98}. NMR experiments\cite{Ohama99} 
reported one V-site above T$_{C}$ (V\( ^{4.5+} \)) but two inequivalent sets of V-sites below T$_{C}$, 
indicating a charge disproportionation of V\( ^{4+} \) ($3d\,^1$, S=$\frac {1}{2}$), V\( ^{5+} \) ($3d\,^0$, S=0). 
These observations suggest that a 
charge ordering (CO) is actually related to the magnetic ordering of V\( ^{4+} \)-V\( ^{4+} \) dimers 
as well as to the lattice distortion, characterized by the wavevector 
\textbf{q}=\((\frac {1}{2}, \frac {1}{2}, \frac{1}{4})\). 
Therefore, $\alpha^\prime$-NaV$_{2}$O$_{5}$ can not be considered as a conventional spin-Peierls 
compound. Up to date, the mechanism of the spin-gap formation and the nature of the CO phase 
transition are still not understood, much effort have been provided to the determination of the 
magnetic ordering, the crystallographic structure and the charge ordering.

After some controversy on the assignment of the high temperature structure space group, 
it is now well accepted that $Pmmn$\cite{Smolinski98} is the best choice for the refinement. 
It describes a centrosymmetric 
crystal structure, of lattice parameters $a$=11.325 \AA, $b$=3.611 \AA~and $c$=4.806 \AA,  
where the only V-site in the unit cell has a mixed-valence state or a static average charge state 
V\( ^{4.5+} \).

Below 34K, a set of weak diffraction peaks appear at \((\frac{h}{2},\frac{k}{2},\frac{\ell}{4}) \) Bragg 
positions with an intensity being roughly 10$^{3}$ times weaker than the Bragg peaks of the \( Pmmn \) 
phase. The refinement of the low temperature (LT) 
crystal structure\cite{Ludecke99} has been greatly hindered by the large 
size of the supercell, $2a\times 2b\times 4c$, including 16 unit cells, and therefore, by the large number 
of atoms to consider. This problem has been circumvented by using a superspace approach, which allows to 
place some constraints between specific atomic parameters. They have shown\cite{Ludecke99} that the 
$Pmmn$ phase remains 
almost unchanged, adjacent unit cells undergoing very small distortions and that the LT space group 
can be {\em reasonably} refined in $Fmm2$, a subgroup of $Pmmn$. They report even still weaker peaks 
obeying a $C$-centering, and therefore violating the $F$-centering, but these reflections have been 
discarded in the refinement due to their very low intensity. Later, de Boer and co-workers arrived at 
an identical conclusion by performing a conventional structure refinement on many measured 
reflections\cite{DeBoer00}. 

$\alpha^\prime$-NaV$_{2}$O$_{5}$ can be seen as a set of layers of two-legs vanadium ladders running 
along the \emph{b}-direction with the rungs along the \emph{a}-direction. V are located inside oxygen 
squared-base pyramids whose corners are shared with neighbor pyramids. The cations Na$^{+}$  are 
located between the layers. The LT distortions, as determined by 
J. L\"{u}decke \emph{et al.}\cite{Ludecke99}, concerns the  pyramids belonging to half of the ladders, 
indeed modulated  and non-modulated ladders alternate along \emph{a}-direction. 
 The displacement highly concerns the {}``bridge oxygen{}'', at the center of the rung, that is shifted 
 0.07\AA\  towards one V of the rung, and alternatively to the other V in the next rungs of the ladder. 
 These latter distortions suggest an alternated localization of the 3$d$ electron on one side of the 
 rung\cite{Mostovoy98} along the ladder. 

The above structure determination has been questioned by experimental results issued from a 
wide variety of techniques. The $^{51}$V NMR study\cite{Ohama99} has determined only two valence states; 
the Raman spectroscopy\cite{Konstantinovic99} measures a number of Raman modes not consistent with the 
number of possible modes in the $Fmm2$ structure. One of the most compelling results is 
probably $^{23}$Na NMR studies\cite{Fagot00,Ohama00} reporting 8 inequivalent Na sites, again in 
complete disagreement with x-ray refinement. Neutron inelastic scattering\cite{Grenier01} and anomalous 
x-ray scattering\cite{Nakao00} results concluded on a zigzag pattern along all the V ladders, but 
only a single-layer model was considered in their calculations. It is interesting to note that these 
2 structural methods 
have converged to a similar solution without using \emph{conventional} crystallography methods, 
whereas \emph{purely} crystallographic refinements are certainly hindered by the inherent 
complexity of the CO in $\alpha^\prime$-NaV$_{2}$O$_{5}$. In this work we present the full charge 
disproportionation on all the ladders and the stacking sequence along the 
\emph{c}-direction in $\alpha^\prime$-NaV$_{2}$O$_{5}$ by means of anomalous x-ray diffraction.

\begin{figure}
\centerline{\epsfxsize=8.5 cm \epsfbox{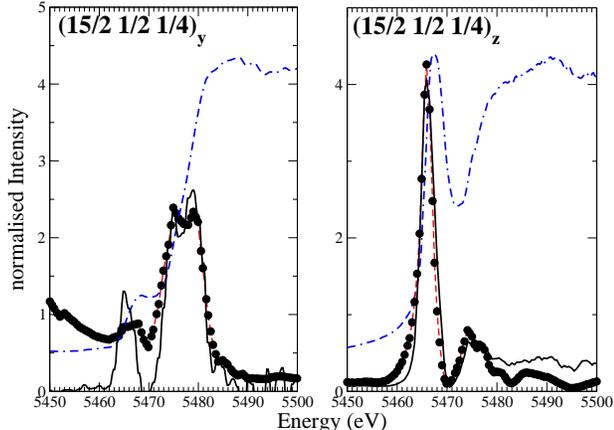}}
\caption{
\( (\frac{15}{2},\frac{1}{2},\frac{1}{4}) \) reflection (\(\bullet\)) and the absorption spectra 
(dot-dash line) taken around the vanadium $K-$edge with the photon polarization along \emph{y} (left panel) 
and along \emph{z} (right panel). The curves have been rescaled for 
clarity. Left panel, the reflection is compared with \(\partial f^{\prime\prime}/\partial E \) 
(solid line). Right 
panel, the reflection is compared (solid line) with one of the two possible
charge ordering models (model 2, Fig. 3). The differences between both polarization states, \emph{y} and
\emph{z}, comes exclusively from the anisotropy of the vanadium site.}
\label{fig_1}
\end{figure}

Anomalous diffraction is a powerful tool in the determination of charge ordering as the X-ray resonant 
scattering amplitude of the atoms is related to the density of occupied and unoccupied electronic 
states\cite{Hodeau01}. By tuning the X-ray energy through an absorption edge of the vanadium, a core 
electron is excited and virtually transits to outer unoccupied states. Below T$_C$, and  as a result 
of the CO, we shall \emph{assume} two new charge states. The anomalous scattering factor for each 
V$^{4+}$ and V$^{5+}$ carries out the information on the charge localization together with  
the {}``bridge{}'' oxygen displacement. 
Under the assumption that the local atomic configuration is not much altered by this charge ordering, 
the resulting near-edge absorption spectra (see Fig. 1) will be slightly shifted towards lower 
(V$^{4+}$) or higher (V$^{5+}$) energies with no significant change in the near-edge structure 
(first 50 eV above the edge). This overall energy shift, amounting a few eV, is known as 
the \emph{chemical shift}. Mathematically speaking, the above sentence implies that 
\( f^{5+}(E+\delta E)\,\approx \,f^{4+}(E)\,\approx \,f^{4.5+}(E+\delta E/2)\). The measured 
absortion spectra corresponds to an average of all the atoms, and therefore \( f(E)=f^{4.5+}(E)\). 
Appropriate spectra for V$^{4+}$ and V$^{5+}$ have been obtained through an adequate choice of $\delta E$.

As the energy dependence of the imaginary part of \emph{\( f(E) \)}, \emph{\( f^{\prime\prime}(E) \)}, 
typically shows a 
discontinuity in a step-like form (see Fig. 1), the chemical shift may yield dramatic anomalies 
if the anomalous factors of the two valence states are subtracted, as it occurs in x-ray diffraction 
experiments for some given, often very weak, reflections. Indeed, the scattering amplitude of the 
valence atom comes 
in multiplied by a phase factor $e^{i\bf {Q}\cdot\bf {r}}$ which, for suitable values of 
\textbf{Q} and \textbf{r}, allows to put atoms with  different valences in opposite phase. 
For such specifically chosen reflections, the vanadium partial structure factor is given 
by \( F\propto f^{4+}(E)+e^{i\pi }f^{5+}(E)\approx f^{5+}(E+\delta E)- f^{5+}(E) \), 
proportional to the \emph{derivative} with respect to energy of the anomalous scattering 
factor, \( F\propto \partial f^{5+}/\partial E \). This is a very powerful property, 
exclusive to anomalous x-ray diffraction, and it has been used for compounds undergoing 
small distortions due to charge and/or orbital ordering\cite{Murakami98,Nakao00}. 

An as-grown $\alpha^\prime$-NaV$_{2}$O$_{5}$ high quality single crystal was used in 
this experiment. It was mounted on the cold finger of a closed cycle He refrigerator, 
and data collection were carried out at the base temperature, 13K.
Experiments were performed at the ID20 beamline in the European Synchrotron Radiation Facility (Grenoble).
An undulator provided the emission of a highly \( \sigma - \)polarized photon flux, the 
monochromatization was performed by a Si-(111) double crystal with an energy resolution 
less than 0.8 eV while mirrors located before and after the monochromator assembly were 
tuned to reject high energy harmonics. The scattering plane was vertical, and therefore 
the incident photon polarization state is \( \sigma \). Measurement of the polarization 
components of the diffracted beam was possible by using the (004) reflection of a pyrolytic 
graphite analyzer crystal. 

A total of 20 different diffraction peaks of the low temperature phase have been measured as a 
function of the incident photon energy, and under different polarization conditions. The energy 
step for all the scan is 0.5 eV. We have found that peaks having a strong anomalous contribution are 
\( (\frac{9}{2},\frac{1}{2},\frac{3}{4}) \), 
\( (\frac{9}{2},\frac{1}{2},\frac{5}{4}) \),
\( (\frac{11}{2},\frac{1}{2},\frac{1}{4}) \), 
\( (\frac{11}{2},\frac{1}{2},\frac{3}{4}) \),
\( (\frac{15}{2},\frac{1}{2},\frac{1}{4}) \) and 
\( (\frac{17}{2},\frac{1}{2},\frac{1}{4}) \) which were measured with the polarization 
along the \emph{b-}direction. In the following we shall use the notation 
\( (h,k,\ell)_{y(z)} \)\cite{Comment1} 
to indicate that the \( (h,k,\ell) \) Bragg intensity has been measured with the polarization 
in the \emph{b(c)-}direction.

Also were measured spectra with the polarization along \emph{c}: 
the \( (\frac{15}{2},\frac{1}{2},\frac{1}{4})_{z} \),  \( (7,0,\frac{1}{2})_{z} \), 
\( (7,1,\frac{1}{2})_{z} \),  \( (6,0,\frac{1}{2})_{z} \), \( (6,1,\frac{1}{2})_{z} \) 
and also, the peaks \( (p,0,0)_{z} \) with \( p=1,3,5,7 \) in \( \sigma -\sigma  \) and 
\( \sigma -\pi  \) geometries. 
The \((\frac{15}{2},\frac{1}{2},\frac{1}{4})_{z} \) peak has a huge anomaly at 5466 eV 
(see Fig. \ref{fig_1}, right panel), as Nakao \emph{et al.}\cite{Nakao00} have also found. This result 
is in contrast with the result for \( (\frac{15}{2},\frac{1}{2},\frac{1}{4})_{y} \), 
where the peak at 5466 eV has completely disappeared (Fig. \ref{fig_1}, left panel), 
thus revealing the phenomenal 
polarization effects this compound exhibits . The other peaks did not show anomalous 
behavior except for the \( (p,0,0)_{z} \) but in the \( \sigma -\pi  \) channel only. 
Indeed \( (p,0,0)_{z} \) peaks are extinct in the high temperature phase because of 
the $n-$glide plane, and one expect signal in the \( \sigma -\pi  \)  channel, alone. 
The lack of anomalous signal in \( \sigma -\sigma  \) channel below T$_C$ in 
the \( (p,0,0)_{z} \) and in the \( (h,k,\frac{1}{2})_{z} \) reflections is a key 
feature that we shall use below. 

\begin{figure}
\centerline{\epsfxsize=7.5 cm \epsfbox{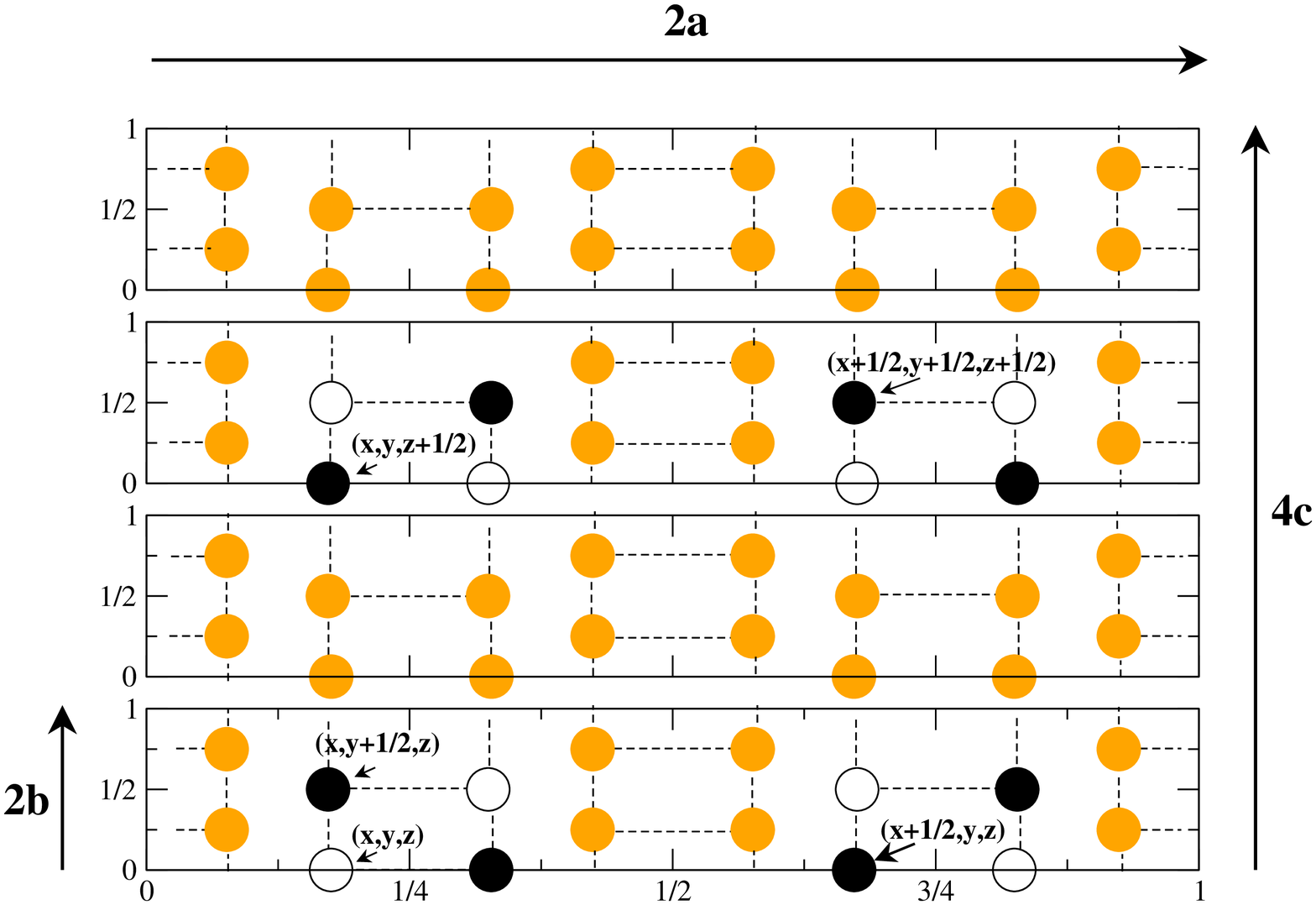}}
\centerline{\epsfxsize=7.5 cm \epsfbox{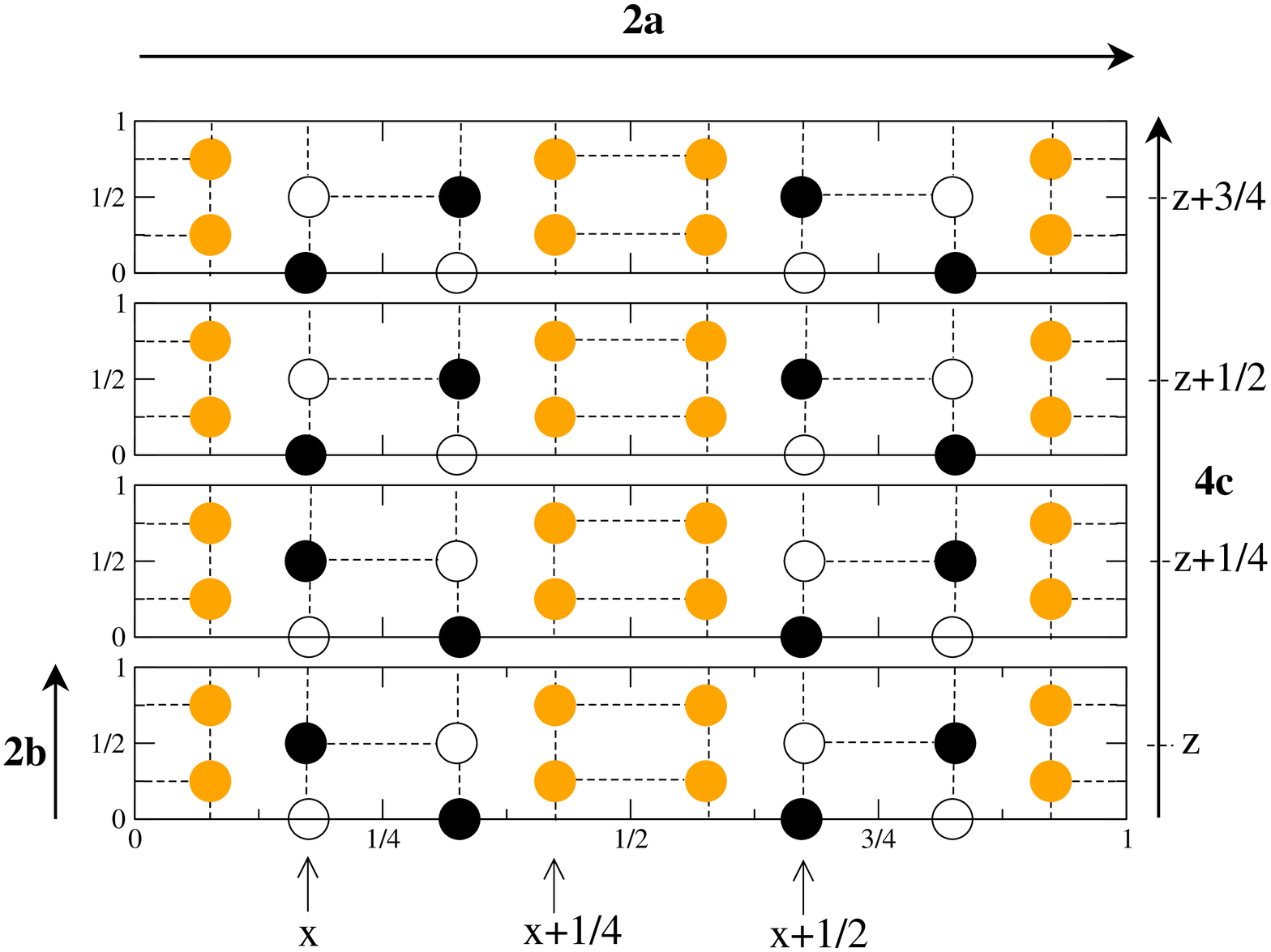}}
\caption{
Schematic representation of vanadium atoms in the 3D structure of $\alpha^\prime$-NaV$_{2}$O$_{5}$. 
V-site assignment based on the use of the \emph{derivative} concept (top panel) for the  
\((\frac{15}{2},\frac{1}{2},\frac{1}{4})\) reflection and of the extinction rule 
found in reflections such as \( (6,0,\frac{1}{2}) \), bottom panel. Black (open) circles are 
V$^{4+}$(V$^{5+}$) and grey circles represent yet unassigned valences}
\label{fig_2}
\end{figure}

Reference fluorescence spectra were taken by removing the analyzer 
crystal and turning the crystal off the Bragg reflection by 
\(1^{\circ } \). The spectra were used  to extract 
adequate values of $f^{\prime}$ and $f^{\prime\prime}$ 
(self-absorption correction has also been carried out) which already include geometrical corrections 
of the tensor form of the scattering factor. 

The reflection \((\frac{15}{2},\frac{1}{2},\frac{1}{4})\) (in all measured polarizations) is of 
very special importance as the energy dependence of the scattered intensity clearly suggests 
a \emph{derivative} effect (Fig. \ref{fig_1}). If two vanadium atoms have the same valence but opposite
phases their contribution would cancel out, whereas if the same atoms have different valence they will
contribute to the \emph{derivative} effect. For \( (\frac{h}{2},\frac{k}{2},\frac{\ell}{4}) \) Bragg peaks, 
adjacent cells along the \emph{a}- and \emph{b}-directions have a phase difference equals to $\pi$. 
In the \emph{c}-direction, the 4 layers have a phase difference equals to 
\( 0 \), \( \frac{\pi }{2} \), \( \pi  \) and \( \frac{3\pi }{2} \),  respectively. In other words, 
an atom in \emph{\( x,y,z \)} is in opposite phase with atoms in \emph{\( x+\frac{1}{2},y,z \)} , 
in \( x,y+\frac{1}{2},z \), in \( x,y,z+\frac{1}{2} \) and in 
\( x+\frac{1}{2},y+\frac{1}{2},z+\frac{1}{2} \). Under this construction, the pair of atoms in 
a single rung has to have different valence state in order to keep the insulating state above T$_C$. 
The first step of the valence attribution is shown in Fig. \ref{fig_2} (upper panel). 
The simple rule found in our spectra readily implies that the V-atoms of different valences 
have to be arranged in zigzag and excludes V$^{4+}$-V$^{4+}$-... chains along the \emph{b}-direction.

\begin{figure}
\centerline{\epsfxsize=6.5 cm \epsfbox{ 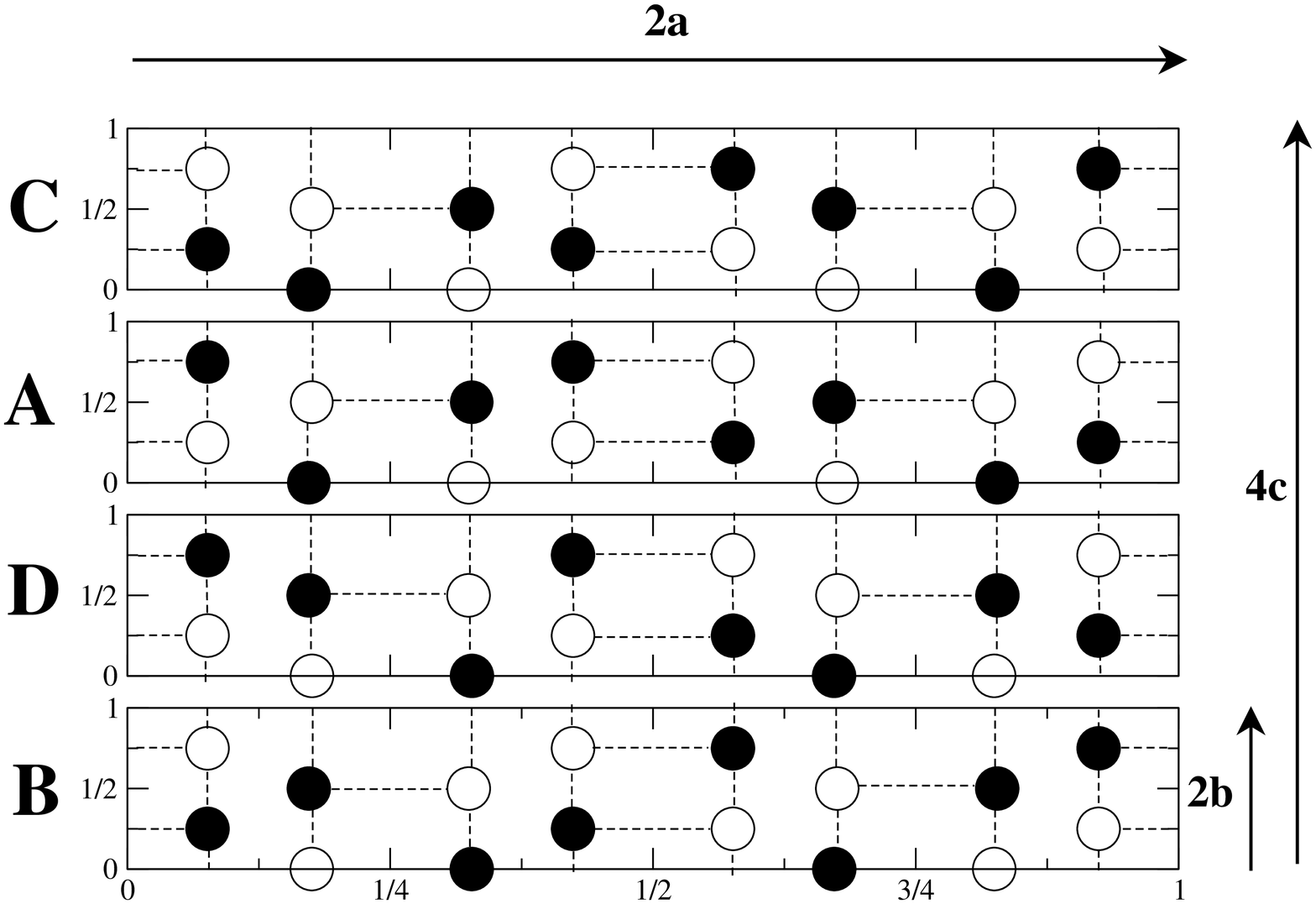}}
\centerline{\epsfxsize=6.5 cm \epsfbox{ 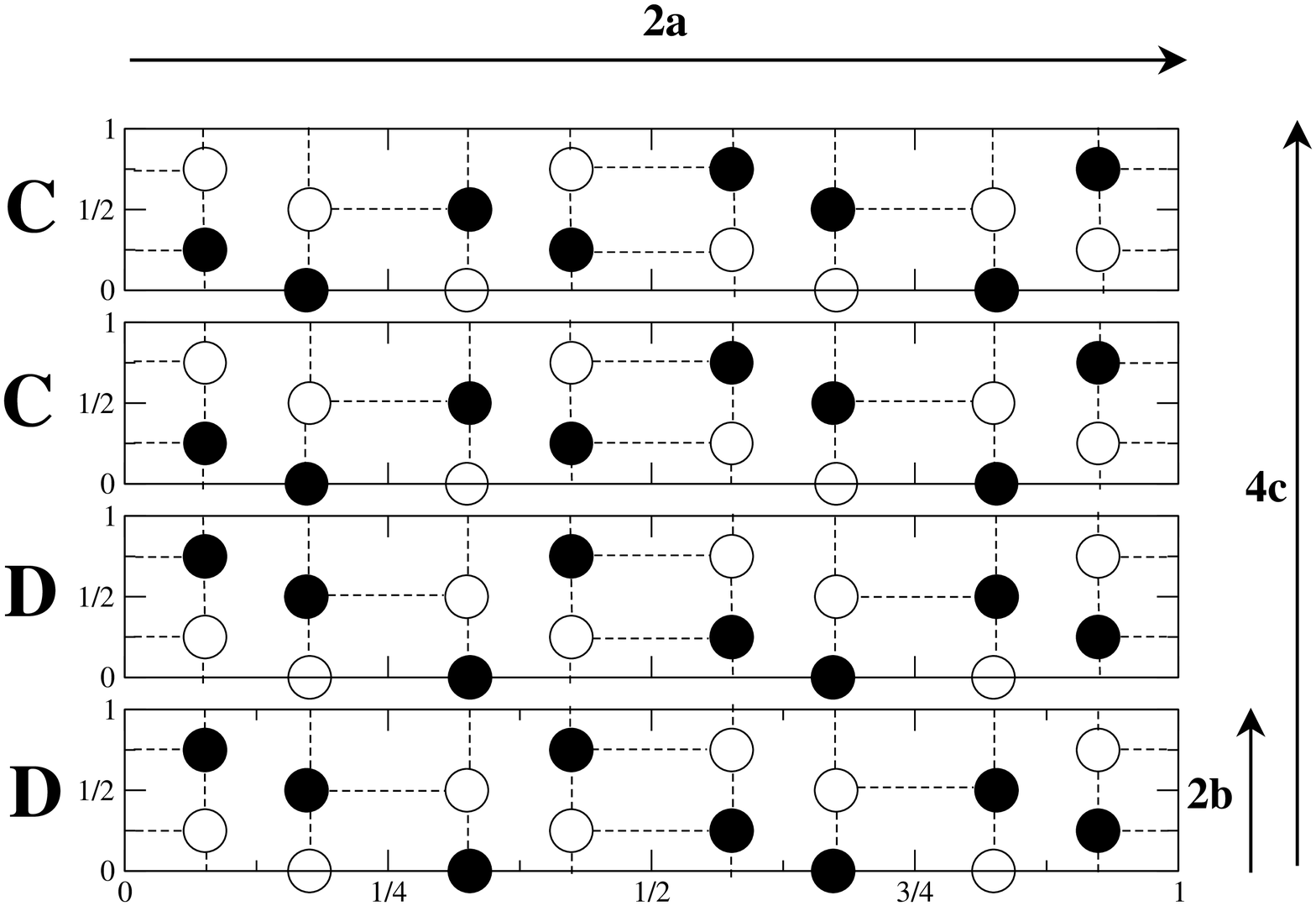}}
\caption{
Possible models of charge orderings resulting from applying the extinction condition for reflection 
of the type \( (p,0,0)\: p\,-odd \) to the results of fig. 2b. Each different layer has been labelled (A, B,
...)  
and we will refer to a particular model by writing up the stacking sequence. Top panel is Model 1, BDAC, 
where each layer has a different arrangement, and appear coupled to the complementary, DBCA (not shown). 
Bottom, Model 2, DDCC, which as in model 1, has to come in with its complementary, BBAA (not shown).}
\label{fig_3}
\end{figure}

For the \( (6,0,\frac{1}{2})_{z} \) peak, no anomalous signal has been detected, and therefore 
all V-contributions to the structure factor have 
to cancel out. For this reflection, atoms on a same \emph{(a,b)}-plane have the same phase, whereas atoms 
in \( z+\frac{1}{4} \) are in opposite phase. Therefore, we infer that the second layer along 
\emph{c}-direction should exhibit the same CO pattern as the first one, and identically between 
layers 3 and 4, as it is shown in Fig. \ref{fig_2} (lower panel).

Finally, an atom at \emph{x} have a phase difference equals to \( \pi  \) with respect to atoms in  
\( x+\frac{1}{4} \) for \( (p,0,0)\: p\:-odd \) peaks. Extinction for these reflections immediately 
implies that ladders centered at \( x=\frac{1}{2} \) have to be modulated as the one in 
\( x=\frac{1}{4} \),  Fig. \ref{fig_3}. By applying the above mentioned conditions, \emph{derivative} in 
\((\frac{15}{2},\frac{1}{2},\frac{1}{4})\) and the absence of intensity in  \( (6,0,\frac{1}{2})_{z} \) 
peak, to the ladders at \( x=\frac{1}{2} \) one get the patterns in Fig. \ref{fig_3}. V-extinction 
considerations in the above set of Bragg reflections  yield CO in all ladders and therefore rules 
out \( Fmm2 \) space group. The symmetry is of $F2$ type with 8 inequivalent sites for Na. 
Note that no fit has been needed so far.

\begin{figure}
\centerline{\epsfxsize=8.5 cm \epsfbox{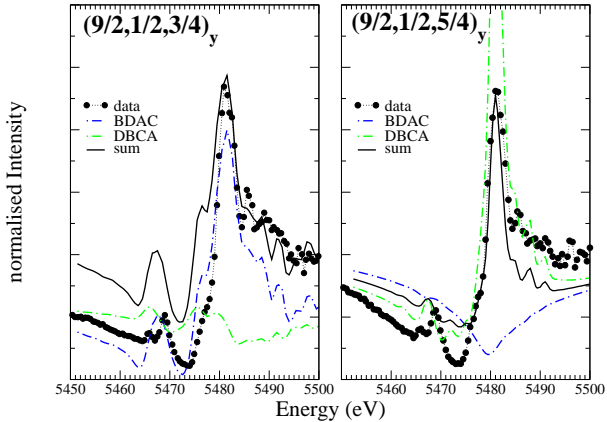}}
\caption{
Experimental data (once corrected from absorption) for peaks \( (\frac{9}{2},\frac{1}{2},\frac{\ell}{4})_y \), 
with $\ell=3$ (left) and  $\ell=5$ (right) and results of simulations with model 1. Model 2 provides
simulations of identical quality.  Whereas the pattern  BDAC reproduces better the $\ell=3$ peak, 
it is the form DBCA for the $\ell=5$ peak. The best result is achieved when both degenerate patterns 
are summed.}
\label{fig_4}
\end{figure}

At this stage of our deductions, we are confronted with the choice among 4
models: indeed the zigzag pattern on one ladder can be shifted along the \emph{b-}direction relatively 
to the neighboring one, as shown in Fig. \ref{fig_3}. Only the ladders centered at \( x=0 \) and in 
\( x=\frac{1}{2} \) (and equally the ladders at \( x=\frac{1}{4} \) and \( x=\frac{3}{4} \)) are related 
to ensure the \emph{derivative} effect. In order to proceed further we shall 
perform a detailed analysis of 3 different reflections : the already studied 
\((\frac{15}{2},\frac{1}{2},\frac{1}{4})_z\) (see Fig. \ref{fig_1})  and 
\((\frac{9}{2},\frac{1}{2},\frac{3}{4})_y\) and \((\frac{9}{2},\frac{1}{2},\frac{5}{4})_y\) 
(see Fig. \ref{fig_4}).

The fit of reflections \((\frac{9}{2},\frac{1}{2},\frac{3}{4})_y\) and 
\((\frac{9}{2},\frac{1}{2},\frac{5}{4})_y\) shows that none of the  BDAC or DBCA patterns of model 1 alone can 
reproduced both reflections, as it is shown in Fig. \ref{fig_4}. The same occurs for the DDCC and 
BBAA patterns of model 2 (not shown). In order to obtain better values of the phases, V-atomic displacements 
along \emph{a} and \emph{c} have been introduced in the fit, but with the constraints 
imposed by the extinction conditions. 
Indeed the sum of patterns BDAC and (the complementary) DBCA of model 1 and, equally, the sum of 
patterns DDCC and AABB of model 2 perfectly accounts for the energy dependence of all measured 
reflections. In order to accommodate these findings, the actual low temperature structure of 
$\alpha^\prime$-NaV$_{2}$O$_{5}$ should contain either domains, related to the stabilization of 
a monoclinic structure, or stacking faults of model 1 or model 2 or of both together. The first 
solution implies the presence of splitted Bragg reflections, which have not been observed either 
in our experiments nor in Van Smaalen's work\cite{VanSmaalen01}, and therefore can be safely ruled out. 
The second solution implies a degenerated ground state which should give rise to very complicated 
phase diagrams. The x-ray diffraction results under pressure (and temperature) by 
Ohwada \emph{et al.}\cite{Ohwada01} nicely show the development of a series of modulation wavevectors 
along the $c^\ast$-direction. This sequence, qualitatively understood within the framework of 
the \emph{Devil's Staircase}-type phase transitions, reflects the presence of competing arrangements 
along \emph{c} of nearly degenerate units, as these proposed in this paper.

As a conclusion, our x-ray anomalous diffraction data has allowed to solve one of the most 
controversial and standing problems in the last years, the charge ordering pattern in 
$\alpha^\prime$-NaV$_{2}$O$_{5}$. We have found that all vanadium ladders are modulated, and different
stacking sequences along the \emph{c-}direction coexist in the structure. 

It is a pleasure to thank J.-P. Boucher, T. Chatterji, O. Cepas and T. Ziman for helpful discussions, 
and S. Blanchard and F. Yakhou for technical help.

\widetext

\begin{references}
\bibitem{Comment1}
All reflections are indexed according to $Pmmn$ setting.
\end{references}
\end{document}